\documentclass[journal=jctcce,manuscript=article]{achemso}

\usepackage{graphicx}
\usepackage{subfigure}
\usepackage{float}
\usepackage{dcolumn}

\usepackage{physics}
\usepackage[version=3]{mhchem}
\usepackage{algorithm}
\usepackage{algpseudocode}
\usepackage{amsmath}
\usepackage{amssymb}
\usepackage{bbm} 
\usepackage{hyperref}
\hypersetup{
    colorlinks,
    linkcolor={blue},
    citecolor={blue},
    urlcolor={black}
}
\usepackage{color}
\usepackage{physics}

\usepackage[dvipsnames]{xcolor}

\newcommand{\eri}[2]{{\left( #1 \middle| #2 \right)}}

\newcommand{\pluseq}{\mathrel{+}=}

\newcommand*{\veck}{{\mathbf{k}}}

\newcommand*{\vecG}{{\mathbf{G}}}
\newcommand*{\vecg}{{\mathbf{g}}}
\newcommand*{\vech}{{\mathbf{h}}}
\newcommand*{\vecF}{{\mathbf{F}}}
\newcommand*{\vecSig}{{\mathbf{\Sigma}}}
\newcommand*{\vecDel}{{\mathbf{\Delta}}}
\newcommand*{\vecJ}{{\mathbf{J}}}

\newcommand*{\vecR}{{\mathbf{R}}}

\newcommand*{\vecI}{{\mathbf{I}}}
\newcommand*{\veczero}{{\mathbf{0}}}

\newcommand*{\emb}{{\mathrm{emb}}}

\newcommand*{\imp}{{\mathrm{imp}}}

\newcommand*{\LO}{{\rm LO}}
\newcommand*{\AO}{{\rm AO}}

\newcommand*{\HF}{{\rm HF}}

\newcommand*{\CC}{{\rm CC}}

\newcommand*{\abinitio}{{\textit{ab initio}\,}}

\makeatletter
\DeclareRobustCommand\onlinecite{\@onlinecite}
\def\@onlinecite#1{\begingroup\let\@cite\NAT@citenum\citealp{#1}\endgroup}
\makeatother

\author{Tianyu Zhu}
\author{Zhi-Hao Cui}
\author{Garnet Kin-Lic Chan}
\email{gkc1000@gmail.com}
\affiliation{Division of Chemistry and Chemical Engineering, California Institute of Technology, Pasadena, CA 91125, USA}

\title{Efficient Formulation of Ab Initio  Quantum Embedding in Periodic Systems: Dynamical Mean-Field Theory}

\begin{tocentry}
\includegraphics[width=\textwidth]{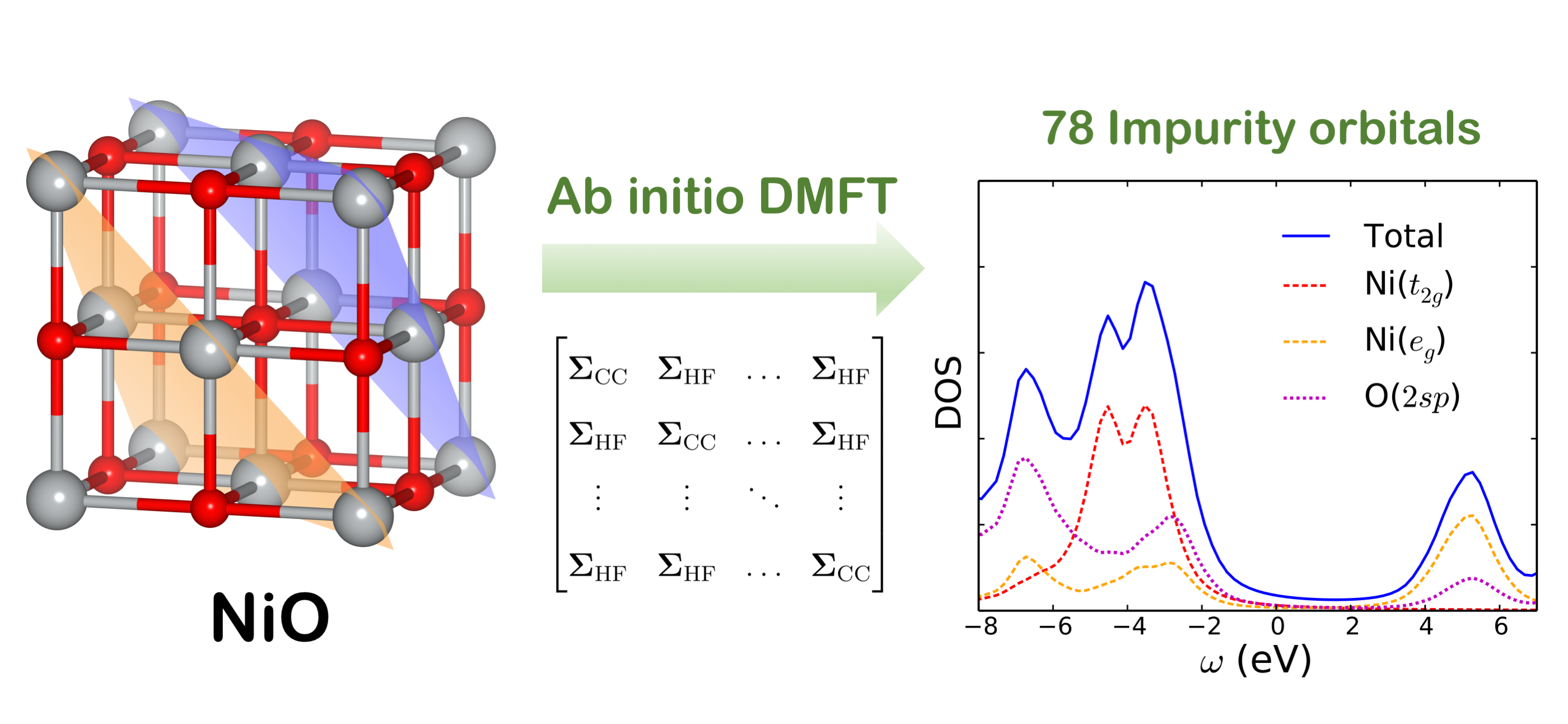}
\end{tocentry}

\begin{document}

\begin{abstract}
We present an efficient \abinitio dynamical mean-field theory (DMFT) implementation for quantitative simulations in solids. Our DMFT scheme employs \abinitio Hamiltonians defined for impurities comprising the full unit cell or a supercell of atoms and for realistic quantum chemical basis sets.
We avoid double counting errors by using Hartree-Fock as the low-level theory. Intrinsic and projected atomic orbitals (IAO+PAO) are chosen as the local embedding basis, facilitating numerical bath truncation. Using an efficient integral transformation and coupled-cluster Green's function (CCGF) impurity solvers, we are able to handle embedded impurity problems  with several hundred orbitals. We apply our \abinitio DMFT approach to study a hexagonal boron nitride monolayer, crystalline silicon, and nickel oxide in the antiferromagnetic phase, with up to 104 and 78 impurity orbitals in spin-restricted and unrestricted cluster DMFT calculations and over 100 bath orbitals. We show that our scheme produces accurate spectral functions compared to both benchmark periodic coupled-cluster computations and experimental spectra.
\end{abstract}

\section{Introduction}\label{sec:intro}

The accurate simulation of strongly correlated electronic materials requires many-body approximations beyond traditional mean-field and
low-order perturbation theories.
An important advance has been the development of dynamical mean-field theory (DMFT), both in its
original single-site formalism~\cite{Georges1992,Georges1996}, as well as in its cluster and multi-orbital extensions~\cite{Lichtenstein2000,Kotliar2001,Hettler2000,Maier2005}. In DMFT, the full interacting solid is mapped onto a local interacting impurity problem, where
the impurity is taken to be a site or cell of local orbitals in the lattice.
The impurity is described by its one-particle Green's function and is self-consistently embedded in an effective non-interacting environment
via a hybridization self-energy. With an appropriate dimensional scaling, DMFT becomes exact in the limit of infinite dimensions~\cite{Metzner1989}
as the only surviving contributions to the self-energy originate from local interactions.
From a quantum chemical perspective, DMFT can also be viewed as a local
correlation theory, because the non-local (inter-cell or inter-site) corrections to the mean-field self-energy are ignored~\cite{Zgid2011}.

Combined with density functional theory (DFT)~\cite{Kohn1965}, DMFT has successfully been used to describe many properties in strongly correlated
$d$ and $f$ electron materials~\cite{Kotliar2006,Held2006,Held2007} including transition metal oxides, heavy fermion systems and high-temperature superconductors.
The DFT+DMFT methodology starts with an \abinitio DFT simulation of the material, which
is then used to construct 
a material-specific low-energy effective Hamiltonian for a small number of localized, strongly interacting, $d$ or $f$
orbital bands. This proceeds via some form of ``downfolding''~\cite{Andersen2000,Werner2010,Werner2016},
where the higher-energy degrees of freedom are approximately integrated out. Commonly, the effective Hamiltonian is
taken to be of generalized Hubbard or Slater-Kanamori form~\cite{Anisimov1991,Kanamori1963} where the one-electron matrix elements
are the matrix elements of the mean-field Hamiltonian in a Wannier basis, and the interaction parameters (such as the Hubbard $U$ parameter)
are obtained empirically or via techniques such as  constrained DFT~\cite{Hybertsen1990,Lichtenstein2001} or the constrained random phase approximation (cRPA)~\cite{Aryasetiawan2004,Aryasetiawan2006,Vaugier2012}.
After the effective Hamiltonian is derived, the corresponding DMFT impurity model is solved using a high-level impurity solver.
This can be a numerically exact
technique, such as continuous-time quantum Monte Carlo (CT-QMC)~\cite{Werner2006,Gull2008,Gull2011} or exact
diagonalization (ED)~\cite{Caffarel1994,Capone2007}, or one of many different approximations, such as configuration
interaction (CI)~\cite{Zgid2011,Zgid2012}, ansatz-based methods~\cite{Bulla2008,White1992,Gutzwiller1963},
or diagrammatic approximations~\cite{Kuramoto1983,Georges1992,Zhu2019}.
The strength of the DFT+DMFT scheme is that the itinerant electrons in the system can often be described well, and in a material specific way, by DFT,
while the DFT hybridization and downfolded interactions yield material specific parameters in the
small, strongly correlated impurity problem that can be treated accurately using the high-level many-body impurity solver.

Yet despite the successes of the DFT+DMFT approach, there are well-known drawbacks and challenges in the
existing formulation and its numerical implementation.
For example, although DFT does not accurately describe the strong correlation between localized $d$ or $f$ electrons,
it nonetheless captures some portion of the
interactions via the Hartree and exchange-correlation functionals. Such Coulomb interactions are also accounted for by DMFT
within the downfolded impurity model, and the combination of the two leads to the prominent issue of double counting of correlations.
This can be corrected for in a variety of ways~\cite{Held2006,Karolak2010,Wang2012,Haule2015}, but much like in other
DFT+wavefunction approaches~\cite{Pollet2002a,Gagliardi2017,Zhu2016,Zhu2019a}, a consistently accurate and rigorously justifiable double counting correction is unknown. 
In addition, there can be numerical issues when downfolding to a low-energy effective Hamiltonian
~\cite{Shinaoka2015,Honerkamp2018}. For example, downfolding to a small number of bands is challenging because
the effective particles are strongly renormalized,  yielding strongly frequency dependent interactions, which is
challenging for impurity solvers; downfolding to 
a large number of bands, on the other hand, requires extracting many parameters, which is difficult to do in a unique and well-conditioned way.
In addition, even for a fixed size of low-energy subspace, 
various choices of Wannier orbitals can lead to substantially different interaction terms.
Finally, the applicability of DMFT in \abinitio simulations of realistic solids remains limited by the computational power of
commonly used impurity solvers. At low temperatures, the CT-QMC solver is usually restricted to a single $d$-shell or $f$-shell impurity, due to the  sign problem and the need for unstable analytic continuation for spectra~\cite{Gull2011,Wang2009}, while ED solvers are restricted to only 4 impurity sites, due to the exponential scaling with impurity size and the need for explicit bath orbitals~\cite{Liebsch2012}.
Approximate methods, for example, based on CI, are naturally of lower cost
~\cite{Zgid2012,lin2013efficient,lu2014efficient}, but even these in practice have only been used to
treat a small number of interacting impurity orbitals~\cite{Go2017}. 

To avoid some of the above issues of DFT+DMFT,  alternative \abinitio DMFT methodologies have been proposed.
Zgid and Chan~\cite{Zgid2011} and Lin \textit{et al.}~\cite{Lin2011}  emphasized the Hartree-Fock (HF)+DMFT approach and further implemented DMFT without using any intermediate downfolded interactions, which might be termed a ``quantum chemical'' approach to DMFT.
The theoretical reason to use  HF+DMFT is that the contributions of the local interactions can be exactly subtracted,
thus completely avoiding double counting, but
the drawback is that non-local interactions (i.e. outside of the impurity) are described only at the HF level.
To address this, more sophisticated diagrammatic methods for including long-range interactions have also been explored.
The first was the GW+DMFT approach~\cite{Sun2002,Biermann2003,Boehnke2016} which starts from a beyond mean-field (e.g. GW)
description of the solid, and subsequently self-consistently embeds the impurity self-energy and polarization propagator. More recently, Zgid and co-workers developed the self-energy embedding theory (SEET)~\cite{Kananenka2015,Lan2015,Rusakov2019},  which is similar to the above, but uses only the embedding of the self-energy. While these developments are promising,
in practice, applications have remained more limited than those with DFT+DMFT and have sometimes retained problematic issues of that approach. For example, GW+DMFT has so far been applied within the downfolded Hamiltonian picture~\cite{Boehnke2016,Choi2016a}, rather than via a fully quantum chemical approach, while
applications using HF+DMFT and SEET to infinite periodic systems, while retaining all interactions, have only appeared at the
level of hydrogen cubes~\cite{Zgid2011} and 1D chains~\cite{Rusakov2019} in a minimal basis.

For quantitative predictions of experimentally relevant properties within the quantum chemical \abinitio approach to DMFT,
it is clear that one requires calculations beyond a few impurity orbitals and small basis sets.
From our perspective,  obstacles to such calculations have been technical rather than
conceptual in nature. One significant obstacle has simply been the complexity of implementing quantum chemistry,  in other words, using realistic  basis sets and computing all long-range matrix elements, in solids.
However, recent years have seen new progress in periodic quantum chemistry infrastructure~\cite{Sun2018,TURBOMOLE},
and the situation is considerably advanced from the time of Ref.~\cite{Zgid2011} and Ref.~\cite{Lin2011}. 
A second obstacle has been the lack of impurity solvers that can handle general (i.e. 4-index) Coulomb matrix elements
for many orbitals.
But similarly, recent implementations of DMRG-based Green's function~\cite{Ronca2017} and coupled-cluster Green's function solvers~\cite{McClain2016,Zhu2019,Shee2019} for quantum chemical Hamiltonians, means that impurity problems with tens to hundreds of interacting orbitals are now practically accessible, depending on the strength of correlation.

In this work, we describe a quantum chemical framework for DMFT that takes advantage of these computational advances
to enable quantitative calculations on non-trivial solids.
Our implementation uses elements of the embedding implementation developed for  density matrix embedding theory (DMET)~\cite{Knizia2012a,Knizia2013c,Wouters2016a}, described in a companion paper~\cite{Cui2019}. 
Based on this framework, we develop a DMFT algorithm that can treat impurities with a large number of orbitals in a realistic basis.
For example, we take our impurity to be
the full unit cell or even a supercell of atoms, in a basis of polarized valence double-zeta quality, yielding in our
largest calculations impurities with over 100 orbitals. 
Several points should be highlighted.
First, following the work of Zgid and Chan~\cite{Zgid2011} and Lin et al.~\cite{Lin2011}, our DMFT scheme starts from the HF approximation, and is thus free of double counting errors.
While non-local interactions remain treated at the mean-field level,  because our large impurities  span multiple atoms, this significantly ameliorates
the neglect of such non-local fluctuations.
Second, in order to handle these large impurity problems, we rely on several techniques. First, we use a coupled-cluster (CC) impurity solver~\cite{Zhu2019,Shee2019} (at the coupled-cluster singles and doubles (CCSD) level) which allows us to treat more than 200 embedded (impurity plus bath) orbitals.
In addition, we reuse the integral transformations described in the accompanying paper~\cite{Cui2019} and the efficient infrastructure of the Python-based
Simulations of Chemistry Framework (\textsc{PySCF})~\cite{Sun2018} to carry out the periodic mean-field and impurity Hamiltonian construction.
Finally, we use real-frequency bath construction and truncation techniques  to systematically converge the DMFT hybridization and spectra.

This paper is organized as follows. In Sec. \ref{sec:theory}, we describe the DMFT formalism and the necessary ingredients in the
computational framework, including the choice of local impurity basis, method of integral transformation and bath construction and truncation.
We also recapitulate the use of CC impurity solvers in DMFT. The computational details are then shown in Sec. \ref{sec:comp}. In Sec. \ref{sec:result}, we present DMFT results on several weakly and strongly correlated materials, including 2D hexagonal boron nitride, 3D bulk silicon, and nickel oxide in the antiferromagnetic (AFM) phase. We finally draw conclusions in Sec. \ref{sec:conclusion}.

\section{Theory}\label{sec:theory}
\subsection{DMFT formalism}\label{sec:dmft}
In this section, we describe the DMFT formalism and the algorithm we use. Throughout this paper, we use the term DMFT to include both single-site
DMFT and the more general cluster DMFT (CDMFT). Both DMFT and CDMFT have been extensively reviewed in the literature~\cite{Maier2005,Kotliar2006,Held2007}, so we only provide a description sufficient for the numerical considerations here. For example, we only describe bath-based DMFT while the bath-orbital-free
DMFT formulations used with CT-QMC solvers are not discussed.

Consider the Hamiltonian of a periodic crystal in an orthogonal basis adapted to the
translational symmetry of a unit cell $\mathcal{C}$ (not necessarily primitive) with $N_\mathcal{C}$ sites,
\begin{equation}\label{eq:Ham}
  \hat{H} = \sum_{pq \in \mathcal{C}} \sum_{\veck} {h}_{pq}(\veck) {a}^\dag_{p\veck} {a}_{q\veck} \\
  + \frac{1}{2} \sum_{pqrs \in \mathcal{C}} \sum_{\veck_p\veck_q\veck_r} 
        {V}_{pqrs}(\veck_p,\veck_q,\veck_r) 
       {a}^\dag_{p\veck_p} {a}^\dag_{q\veck_q} {a}_{r\veck_r} {a}_{s\veck_p+\veck_q-\veck_r} ,
\end{equation}
where $h_{pq}$ and $V_{pqrs}$ are the one- and two-particle interaction matrix elements
of the basis functions, indexed by labels in the unit cell $\mathcal{C}$,
and momentum conservation is modulo lattice vectors. 
The non-interacting lattice Green's function is block-diagonal in $\veck$ space and can be written as
\begin{equation}\label{eq:g0}
\vecg (\veck,\omega) = [(\omega + \mu) \vecI -\vech(\veck)]^{-1},
\end{equation}
where $\mu$ is the chemical potential. 
The interacting lattice Green's function is related to the non-interacting Green's function via  Dyson's equation:
\begin{equation}\label{eq:Gk}
\vecG (\veck,\omega) = [\vecg^{-1} (\veck,\omega)- \vecSig(\veck, \omega)]^{-1} = [(\omega + \mu) \vecI -\vech(\veck)  -\vecSig(\veck, \omega)]^{-1} ,
\end{equation}
where the self-energy $\vecSig(\veck, \omega)$ accounts for the many-body correlation effects. The lattice Green's function in real space can then be obtained by a Fourier transform:
\begin{equation}\label{eq:GC}
\vecG (\mathbf{R}=\mathbf{0},\omega) = \frac{1}{N_{\veck}} \sum_{\veck} \vecG (\veck,\omega),
\end{equation}
and the local spectral function is defined as
\begin{equation}
\mathbf{A} (\mathbf{R}=\mathbf{0},\omega) = -\frac{1}{\pi} \Im \vecG (\mathbf{R}=\mathbf{0},\omega+i0^+).
\end{equation}

Directly computing the interacting lattice Green's function in the thermodynamic limit using an accurate many-body method is very expensive.
Quantum embedding methods such as DMFT provide another route to this quantity. In DMFT, the periodic Hamiltonian in Eq.~\ref{eq:Ham} is mapped onto an effective real-space impurity problem, where the unit cell $\mathcal{C}$ (impurity) is embedded in a non-interacting environment. We will discuss the detailed form of the embedding Hamiltonian in the next section, and continue our focus on the general DMFT formalism here. 

The key approximation in DMFT is to replace the $\veck$-dependent lattice self-energy $\vecSig(\veck, \omega)$ with a local self-energy $\vecSig_{\imp}(\omega)$ defined for the $N_\mathcal{C}$ impurity sites:
\begin{equation}
\vecSig(\veck, \omega) = \vecSig_{\imp}(\omega).
\end{equation}
Neglecting $\veck$-dependence is equivalent to ignoring the off-diagonal self-energy between unit cells in real-space, which introduces a local correlation approximation. Given $\vecSig_{\imp}(\omega)$, the lattice Green's function in Eq.~\ref{eq:GC} is approximated as
\begin{equation}\label{eq:GCDMFT}
\vecG (\mathbf{R}=\mathbf{0},\omega) = \frac{1}{N_{\veck}} \sum_{\veck} [(\omega + \mu) \vecI -\vech(\veck)  -\vecSig_\imp(\omega)]^{-1} .
\end{equation}

To determine $\vecSig_{\imp}(\omega)$ we solve a many-body problem for an embedded impurity, which is much simpler than
the many-body problem for the
the whole crystal. In the impurity problem,
to describe the delocalization effects from the impurity-environment interaction, the hybridization self-energy is introduced:
\begin{equation}\label{eq:hyb}
\vecDel (\omega) = (\omega + \mu) \vecI -\vech_\imp  -\vecSig_{\imp}(\omega) - \vecG^{-1}(\mathbf{R}=\mathbf{0},\omega) ,
\end{equation}
where $\vech_\imp$ is the impurity one-particle Hamiltonian. When wavefunction-based impurity solvers (for example, CC in this work) are employed,
the hybridization $\vecDel (\omega)$ is mapped onto couplings between the impurity orbitals and a finite set of non-interacting bath sites. These
interactions are included in the embedding Hamiltonian (see next section) which includes both the effects of local two-particle
interactions, and delocalization into the environment. From the embedding Hamiltonian, the impurity Green's function $\vecG_{\imp} (\omega)$
is computed by the many-body impurity solver, and the corresponding impurity self-energy $\vecSig_{\imp}(\omega)$ is defined as
\begin{equation}\label{eq:sigmaimp}
\vecSig_{\imp}(\omega) = (\omega + \mu) \vecI -\vech_\imp -\vecDel(\omega) - \vecG_{\imp}^{-1} (\omega).
\end{equation}

Since the impurity self-energy $\vecSig_{\imp}(\omega)$ and hybridization 
 $\vecDel (\omega)$ are defined in terms of each other, the DMFT equations are solved self-consistently 
until the impurity Green's function $\vecG_{\imp} (\omega)$ and the lattice Green's function $\vecG (\mathbf{R}=\mathbf{0},\omega)$ agree:
\begin{equation}
\vecG_{\imp} (\omega) = \vecG(\mathbf{R}=\mathbf{0},\omega).
\end{equation}
In practice, convergence of the self-consistency can be monitored also in terms of $\vecSig_{\imp}(\omega)$
or the hybridization  $\vecDel (\omega)$.
In this work, we assess convergence from the change of  $\vecDel (\omega)$ between DMFT iterations, and use the DIIS technique~\cite{Pulay1980}
to accelerate convergence.

\subsection{Embedding Hamiltonian}\label{sec:Hemb}

We next describe how to construct the embedding Hamiltonian in DMFT. We take
as our starting point the \abinitio quantum embedding implementation for DMET discussed in Ref.~\cite{Cui2019}, where several important computational choices
and algorithms are discussed, including (1) defining the lattice and impurity basis starting from a mean-field calculation; (2) the choice of local orthogonal basis; (3) efficient integral transformation algorithms. We refer to Ref.~\cite{Cui2019} for additional details.

\paragraph{HF-based DMFT}

In Sec.~\ref{sec:dmft}, the DMFT formalism used the bare one-particle Hamiltonian in the definition of the non-interacting lattice
Green's function (Eq.~\ref{eq:g0}). This is a reasonable choice for lattice models with purely local interactions,
but in realistic solids, long-range Coulomb interactions play an important role. In this case,  mean-field methods such as DFT and HF can be
employed to provide the non-local self-energy between unit cells in DMFT. 

In this work, we will only use Hartree-Fock non-local self-energies to avoid
the double counting problems that arise when using DFT self-energies.
The HF Green's function is defined as
\begin{equation}\label{eq:gHF}
\vecg_{\HF} (\veck,\omega) = [(\omega + \mu) \vecI -\vecF(\veck)]^{-1},
\end{equation}
where $\vecF(\veck) = \vech(\veck) + \vecSig_{\HF}(\veck)$ is the lattice Fock matrix that includes the static HF self-energy $\vecSig_{\HF}(\veck)$.
The HF self-energy may be computed from a self-consistent HF solution of the crystal, or from a non-self-consistent density
matrix and orbitals (such as DFT orbitals). Since the self-energy is diagrammatically defined in both cases,
double-counting can be completely avoided.

The one-particle Hamiltonians in both the crystal and embedding problem are then replaced by the Fock matrix:
\begin{equation}
\vech(\veck) \rightarrow \vecF(\veck),~ \vech_\imp \rightarrow \vecF_\imp .
\end{equation}
For example, the lattice Green's function $\vecG(\mathbf{R}=\mathbf{0},\omega)$ is now:
\begin{equation}\label{eq:GCHF}
\vecG (\mathbf{R}=\mathbf{0},\omega) = \frac{1}{N_{\veck}} \sum_{\veck} [(\omega + \mu) \vecI -\vecF(\veck)  -\vecSig_\imp(\omega)]^{-1}.
\end{equation}
Accordingly, the definition of the impurity self-energy $\vecSig_{\imp}(\omega)$ also changes:
\begin{equation}\label{eq:sigmaimpHF}
\vecSig_{\imp}(\omega) = (\omega + \mu) \vecI -\vecF_\imp -\vecDel(\omega) - \vecG_{\imp}^{-1} (\omega) .
\end{equation}
In this work, where we use CC as the impurity solver (see below), this implies that
 $\vecSig_{\imp}(\omega)$ is taken as the difference between the CC self-energy and HF self-energy of the impurity:
\begin{equation}
\vecSig_{\imp}(\omega) = \vecSig_{\imp}^{\CC}(\omega) - \vecSig_{\imp}^{\HF} .
\end{equation}
One subtlety to consider is the density matrix to use when computing $\vecSig_{\imp}^{\HF}$.
In a charge self-consistent DMFT calculation, one updates the lattice Fock matrix and impurity
Hartree-Fock self-energy using correlated lattice and impurity density matrices. In this work, we use the initial mean-field impurity density matrix from lattice HF or DFT calculations to build $\vecSig_{\imp}^{\HF}$ and keep $\vecSig_{\imp}^{\HF}$ fixed during the DMFT iterations. We leave
  a charge self-consistent implementation to future work.

\paragraph{Local orthogonal orbital basis}
Until now, we have assumed an orthogonal basis in the formalism. However, in our mean-field calculations we use
crystal Gaussian atomic orbitals (AO), which are not orthogonal. While it is possible
to formulate DMFT with Green's functions and matrix elements in a non-orthogonal basis, the hybridization $\vecDel (\omega)$
acquires a non-trivial high frequency tail that can be hard to represent in a bath formulation~\cite{Kotliar2001,Zgid2011}.
Therefore, it is desirable to work with an orthogonal atom-centered (crystal) local orbital (LO) basis.

As discussed in Ref.~\cite{Cui2019} for \abinitio DMET, we will employ intrinsic (crystal) atomic orbitals and projected (crystal) atomic orbitals (IAO+PAO)~\cite{Knizia2013d,Wouters2016a} as the local orthogonal basis.
IAOs are a set of valence atomic-like orbitals that exactly span the occupied space of the mean-field calculations, and whose construction only requires
projecting the HF/DFT orbitals onto pre-defined valence (minimal) AOs. PAOs, on the other hand, provide the remaining high-energy virtual atomic-like orbitals that are orthogonal to the IAO space. The main advantage of the IAO+PAO scheme is that no numerical optimization is needed to construct
the LOs and the mean-field valence and virtual spaces are explicitly separated.

From the IAO+PAO projections, a transformation matrix from crystal AOs to crystal LOs at each $\veck$ point $C^{\veck,\AO,\LO}$ can be obtained. The real-space one-particle Hamiltonian, Fock matrix and density matrix in the impurity LO basis are computed as:
\begin{equation}\label{eq:h1eimp}
\vech_\imp = \frac{1}{N_\veck} \sum_{\veck} \vech^\LO(\veck)  =  \frac{1}{N_\veck} \sum_{\veck} {C^{\veck,\AO,\LO}}^{\dag} \vech^{\AO}(\veck) C^{\veck,\AO,\LO} ,
\end{equation}
\begin{equation}
\vecF_\imp = \frac{1}{N_\veck} \sum_{\veck} \vecF^\LO(\veck)  =  \frac{1}{N_\veck} \sum_{\veck} {C^{\veck,\AO,\LO}}^{\dag} \vecF^{\AO}(\veck) C^{\veck,\AO,\LO} ,
\end{equation}
\begin{equation}\label{eq:gammaimp}
\gamma_\imp = \frac{1}{N_\veck} \sum_{\veck} \gamma^\LO(\veck)  =  \frac{1}{N_\veck} \sum_{\veck} {C^{\veck,\AO,\LO}}^\dag \mathbf{S}(\veck) \gamma^{\AO}(\veck) \mathbf{S}(\veck) {C^{\veck,\AO,\LO}} ,
\end{equation}
where $\gamma^{\AO}(\veck)$ and $\mathbf{S}(\veck)$ are the one-particle density matrix and overlap matrix in $\veck$ space.

The impurity Hamiltonian can then be formulated as:
\begin{equation}\label{eq:Himp}
  \hat{H}_\imp = \sum_{ij \in \imp} {\tilde{F}}_{ij} {a}^\dag_{i} {a}_{j} \\
  + \frac{1}{2} \sum_{ijkl \in \imp} \eri{ij}{kl} {a}^\dag_{i} {a}^\dag_{k} {a}_{l} {a}_{j} ,
\end{equation}
where $i,j,k,l$ stand for impurity local orbitals and $\eri{ij}{kl}$ denotes a two-electron repulsion integral (ERI). To avoid double counting in DMFT,
the Hartree-Fock contribution to the impurity self-energy needs to be removed from the one-particle impurity Hamiltonian:
\begin{equation}
\tilde{F}_{ij} = (F_\imp)_{ij} - \sum_{kl \in \imp} (\gamma_\imp)_{kl} [\eri{ij}{lk} - \frac{1}{2} \eri{ik}{lj}] .
\end{equation}
This subtraction eliminates double counting in HF-based DMFT.

\paragraph{ERI transformation}
The most expensive step in forming the impurity Hamiltonian is to construct the impurity ERIs $\eri{ij}{kl}$. In our implementation, we start
from the $\veck$-space Gaussian density fitting (GDF)~\cite{Whitten73,Sun17} AO integrals $\eri{L\veck_L}{p\veck_p q\veck_q}$, where $L$ is a crystal Gaussian auxiliary basis. Note that $\veck_L$, $\veck_p$ and $\veck_q$ satisfy  momentum conservation modulo lattice vectors: $\veck_L = \veck_p - \veck_q$. The detailed ERI transformation algorithm is presented in Algorithm~\ref{alg:eri with gdf}. This algorithm scales as $\mathcal{O}\qty(n_{\veck}^2 n_{L} n^3_{\AO})$ for the $\veck$-AO to $\veck$-LO transformation step and $\mathcal{O}\qty(n_{\veck} n_{L} n^4_{\AO})$ for the ERI contraction step.
This should be compared to the steep  $\mathcal{O}\qty(n_{\veck}^3 n^5_{\AO})$ scaling of the transformation
  without density fitting techniques.
We note that this algorithm is the same as would be used for the ERI transformation in the non-interacting bath formalism of DMET. 

\begin{algorithm}[hbt]
  \caption{Pseudocode for the impurity ERI transformation in DMFT using GDF.} 
  \label{alg:eri with gdf}
  \begin{algorithmic}[1]
    \For{all $\veck_{L}$}
    \For{$\qty(\veck_{p}, \veck_{q})$ that conserves momentum} 
    \State Transform $\eri{L}{p\veck_p q\veck_q}$ to $\eri{L}{i\veck_p j \veck_q}$ by $C^{\veck, \AO, \LO}$
    \Comment{$\veck$-AO to $\veck$-LO}
    \State $\eri{L}{\veczero i \veczero j} \pluseq \frac{1}{N_{\veck}} \eri{L}{i \veck_p j \veck_q}$ \Comment{FT to the reference cell $\vecR = \veczero$}
    \EndFor
    \State $\eri{ij}{kl} \pluseq \frac{1}{N_{\veck}} \sum_{L} \eri{\veczero i \veczero j}{L} \eri{L}{\veczero k \veczero l}$
    \Comment{ERI contraction}
    \EndFor
  \end{algorithmic}
\end{algorithm}

\paragraph{Bath truncation} In bath-based DMFT, the hybridization $\vecDel(\omega)$ is represented by a finite set of discrete bath sites and couplings. In this work, we choose to approximate $\vecDel(\omega)$ along the real frequency axis~\cite{lu2014efficient,Bulla2008,Peters2011,Ganahl2015}
so that dynamical quantities (e.g. spectral functions) can be computed more accurately than when fitting along the imaginary frequency axis~\cite{Liebsch2012,Koch2008,wolf2015imaginary}. We consider $\vecDel(\omega)$ as the Hilbert transform~\cite{DeVega2015}
\begin{equation}
\vecDel(\omega) = \int \frac{\vecJ(\epsilon)}{\omega - \epsilon} d\epsilon 
\end{equation}
with the spectral density 
\begin{equation}
\vecJ(\epsilon) = -\frac{1}{\pi} \mathbf{Im} \vecDel (\epsilon + i\eta) ,
\end{equation}
where $\eta$ is a broadening parameter. The Hilbert transform integral can be approximated by a numerical quadrature (Gauss-Legendre in this work) along the real axis
\begin{equation}\label{eq:bathquad}
\vecDel(\omega) = \sum_{n=1}^{N_\omega} w_n \frac{\vecJ(\epsilon_n)}{\omega - \epsilon_n} ,
\end{equation}
where $w_n$ and $\epsilon_n$ are the weights and positions of the $N_\omega$ quadrature grid points. To derive the couplings between the
impurity and bath sites, we diagonalize the spectral density
\begin{equation}
\vecJ(\epsilon_n) = \mathbf{U}^{(n)} \mathbf{\lambda}^{(n)} {\mathbf{U}^{(n)}}^\dag.
\end{equation}
Eq.~\ref{eq:bathquad} then becomes:
\begin{equation}\label{eq:bathfit}
\Delta_{ij}(\omega) = \sum_{n=1}^{N_\omega} \sum_{k=1}^{N_\mathcal{C}} \frac{V_{ik}^{(n)} V_{jk}^{(n)}}{\omega - \epsilon_n} ,
\end{equation}
with
\begin{equation}\label{eq:coupling}
V_{ik}^{(n)} = w_n^{\frac{1}{2}} U_{ik}^{(n)} (\lambda_{kk}^{(n)})^{\frac{1}{2}} .
\end{equation}
Thus, $V_{ik}^{(n)}$ and $\epsilon_n$  can be interpreted as the impurity-bath couplings and energy levels of bath orbitals. With this bath discretization, we can finally define the full embedding Hamiltonian
\begin{equation}\label{eq:Hemb}
  \hat{H}_\emb = \hat{H}_\imp + \sum_{n=1}^{N_\omega} \sum_{k=1}^{N_\mathcal{C}} \Big( \sum_i V_{ik}^{(n)} (a_i^\dag a_{nk} + a_{nk}^\dag a_i) + \epsilon_n a_{nk}^\dag a_{nk} \Big) .
\end{equation}

It is known that bath discretization introduces discretization errors into DMFT, thus many bath orbitals per impurity site are required to minimize this error. In our case, the number of bath orbitals is formally $N_b = N_\omega N_\mathcal{C}$, which can easily be as many as a few hundred, as $N_\mathcal{C}$ includes all orbitals in the unit cell. To reduce the bath size and thus computational cost,
we employ several strategies to truncate the bath degrees of freedom, while minimizing the error introduced.
First, we only couple bath orbitals to the IAOs (valence orbitals). Note that PAOs (non-valence virtuals) are still included in the embedding problem and interact with the IAOs. With this choice, the number of bath orbitals is reduced to $N_b = N_\omega N_\mathrm{IAO}$. Second, we remove the bath orbitals which are very weakly coupled to the impurity. As seen in Eq.~\ref{eq:coupling}, the scale of the bath coupling is set by the eigenvalues $\lambda_{kk}^{(n)}$.
By dropping the bath orbitals with eigenvalues below a threshold, we can further decrease the bath dimension as necessary.

We have now described all the ingredients needed to perform \abinitio DMFT calculations in a practical and efficient manner in realistic solids. For clarity, we
summarize the detailed steps in Algorithm~\ref{alg:dmft}. Note that the chemical potential $\mu$ is adjusted to ensure the electron count of the impurity is  correct (in these calculations, since multiples of the full unit cell are used as the impurity and there is no doping,
  the impurity electron count is an integer).

\begin{algorithm}[hbt]
  \caption{Pseudocode for \abinitio DMFT in realistic solids.} 
  \label{alg:dmft}
  \begin{algorithmic}[1]
  \State IAO+PAO construction
  \State Integral transformation: Eqs.~\ref{eq:h1eimp}-\ref{eq:gammaimp} and Algorithm~\ref{alg:eri with gdf}
  \State \textbf{Input:} $N_e$ (electron number in $\mathcal{C}$), $N_\omega$ and quadrature range
  \While{$|\mathbf{Tr} \gamma_\imp^{\mathrm{CC}} - N_e| > \delta_e$}
  \State Choose chemical potential $\mu$ 
  \State Initial guess of impurity self-energy $\vecSig_\imp(\omega)$
  \State Initialize hybridization $\vecDel(\omega)$ via Eq.~\ref{eq:hyb}
  \Comment HF hybridization if $\vecSig_\imp(\omega) = \mathbf{0}$
    \While{ $||\vecDel_{i+1}(\omega) - \vecDel_{i}(\omega)|| > \theta $}
    \State Discretize $\vecDel(\omega)$ via Eq.~\ref{eq:bathfit}
    \State Form $\hat{H}_\emb$ via Eq.~\ref{eq:Hemb}
    \State Solve embedding problem with HF at fixed $\mu$
    \Comment Electron number can change 
    \State Compute $\vecG_\imp(\omega)$ using CCSD-GF solver
    \State Calculate $\vecSig_\imp(\omega)$ via Eq.~\ref{eq:sigmaimpHF}
    \State Calculate $\vecDel(\omega)$ via Eq.~\ref{eq:hyb} and update using DIIS
    \EndWhile
  \EndWhile
  \State Compute $\vecG(\mathbf{R}=\mathbf{0},\omega)$ over desired frequencies via Eq.~\ref{eq:GCHF}
  \end{algorithmic}
\end{algorithm}

\subsection{Coupled-cluster impurity solvers}
In this work, we will use a coupled-cluster (CC) impurity solver to compute $\vecG_\imp(\omega)$ from the embedding Hamiltonian in Eq.~\ref{eq:Hemb}.
We recently studied the coupled-cluster Green's function (CCGF) approximation as an impurity solver
in DMFT~\cite{Zhu2019} (see also Ref.~\cite{Shee2019}) and showed that it performed well for small impurity clusters in Hubbard models.
Here, we further explore its capabilities in the \abinitio setting.

A detailed presentation of the CCGF formalism can be found in Refs.~\cite{Zhu2019,Nooijen1992,Nooijen1993,McClain2016,Bhaskaran-Nair2016,Furukawa2018}. Here, we will only comment on a few points related to DMFT. First, we use the coupled-cluster singles and doubles Green's function approximation (CCSD-GF)
as the impurity solver here.  The cost of ground-state CCSD scales as $\mathcal{O}(N_\emb^6)$ and CCSD-GF scales as $\mathcal{O}(N_\omega N_{\mathcal{C}} N_\emb^5)$, where $N_\emb$ is the total number of orbitals (sites) in the embedding problem and $N_{\mathcal{C}}$ is the number of impurity orbitals.
In practice, the CCSD-GF calculation can be parallelized over $N_\omega$ and $N_{\mathcal{C}}$. This low cost compared to ED
allows us to treat around 200 embedding orbitals in DMFT. Second, the CCSD-GF is computed directly on the real frequency axis with a broadening factor $\eta$. We find that in \abinitio calculations, when $\eta$ is small ($<$ 0.5 eV), causality issues may exist where the imaginary part of the CCSD self-energy is not always negative (see Fig. S2 for an example). We refer readers to Ref.~\cite{Zhu2019} for a practical solution, where one computes the impurity plus bath self-energy and then uses the impurity block of the self-energy matrix, instead of directly computing the impurity self-energy matrix. This solution increases the CCSD-GF cost scaling slightly to
$\mathcal{O}(N_\omega N_{\emb} N_\emb^5)$. We use this more expensive technique in the final calculation of spectra after DMFT convergence.

\section{Computational Details}\label{sec:comp}

We applied the above DMFT scheme to study three realistic solids: weakly correlated 2D hexagonal boron nitride (h-BN) and crystalline silicon (Si),
and a prototypical ``strongly correlated'' problem, nickel oxide (NiO) in the AFM-II phase. The experimental lattice constants were used for all calculations: 2.50\AA~for h-BN monolayer~\cite{Li11hBN}, 5.417\AA~for Si~\cite{Tobbens01Si} and 4.17\AA~for NiO~\cite{Cheetham83NiO}. For h-BN, we used a vacuum
spacing of 20\AA~along the $z$ axis to avoid image interactions between neighboring sheets. In h-BN and Si we used (super)cells of the primitive cell
as the impurity. In NiO, we used a supercell with two Ni and two O atoms along the [111] direction to allow for AFM spin symmetry breaking within the cell.

All mean-field calculations were performed and all integrals were generated using the \textsc{PySCF} quantum chemistry package~\cite{Sun2018}. Norm-conserving GTH-PADE pseudopotentials~\cite{Goedecker96,Hartwigsen98} were employed to replace the core electrons. The GTH-DZVP basis set was used for h-BN and Si, while the GTH-DZVP-MOLOPT-SR basis set~\cite{VandeVondele2007} was used for NiO. This corresponds to 26, 26 and 78 AOs in the impurity unit cells of h-BN, Si and NiO, and 104 AOs in our largest impurity supercell of Si. The minimal basis sets GTH-SZV (h-BN and Si) and GTH-SZV-MOLOPT-SR (NiO) were used as the pre-defined AOs to construct the IAOs, leading to 8, 8 and 28 valence orbitals in each unit cell respectively. Even-tempered auxiliary Gaussian basis sets were used to compute the GDF integrals. Uniform $6\times6\times1$ (h-BN), $4\times4\times4$ (Si) and $4\times4\times4$ (NiO) $\veck$-point meshes were adopted for
the mean-field and DMFT calculations. All meshes were $\mathbf{\Gamma}$-centered.

Unless otherwise specified, we used HF as the initial lattice mean-field in the DMFT embedding. The CCSD-GF method was implemented based on the
CCSD and EOM-CCSD routines from the \textsc{PySCF} package, and the DMFT algorithm was implemented in the \textsc{Potato} module.
A spin-restricted CCSD-GF (RCCSD-GF) solver was used for h-BN and Si, while a spin-unrestricted CCSD-GF (UCCSD-GF) solver within the spin-unrestricted DMFT formalism was employed for NiO. A simplified and flexible variant of the GCROT method (GCROT($m,k$))~\cite{DeSturler1999} was used to solve the
CCSD-GF linear response equations. Gauss-Legendre quadrature was used on frequency intervals of $[-1.0+\mu, 1.0+\mu]$ a.u. (h-BN), $[-0.6+\mu, 1.0+\mu]$ a.u. (Si) and $[-0.5+\mu, 0.5+\mu]$ a.u. (NiO) when discretizing the hybridization. We used a broadening factor of $\eta=0.1$ a.u. during the DMFT self-consistent cycles and switched to a smaller $\eta$ (depending on the required resolution) in the final production runs to compute $\vecG(\mathbf{R}=\mathbf{0},\omega)$. The DMFT self-consistency was converged to $||\vecDel_{i+1}(\omega) - \vecDel_{i}(\omega)|| < \theta = 10^{-4}$ a.u. between two DMFT iterations.

\section{Results and Discussion}\label{sec:result}
\subsection{2D h-BN}

We first investigate the performance of our \abinitio DMFT scheme for the 2D h-BN monolayer. We chose the impurity to be an h-BN unit cell,
including the $2s2p3s3p3d$ orbitals for both boron and nitrogen atoms (26 impurity orbitals in total). The corresponding IAOs are
projected $2s2p$ orbitals, giving 8 IAOs coupled to the bath. To study the convergence of DMFT with respect to the number of bath orbitals, we used a series of Gauss-Legendre quadratures to discretize the hybridization: $N_\omega=4,8,12,16$. This led to a total number of $N_b=32,64,96,128$ bath orbitals. 

\begin{table}[hbt]
	\centering
	\caption{Direct and indirect band gaps (in eV) of 2D h-BN.}
	\label{tab:BN_gap}
	\begin{tabular}{cccc}
	\hline\hline
Method	 &  $\mathbf{K} \rightarrow \mathbf{K}$  &  $\mathbf{K} \rightarrow \mathbf{\Gamma}$ &  $\mathbf{\Gamma} \rightarrow \mathbf{\Gamma}$  \\
	\hline
HF   &   11.31  &  10.70   &  13.14   \\
PBE &  4.61 & 5.90 & 7.37  \\
DMFT(i26,b32) & 5.69 & 6.85 & 10.10 \\
DMFT(i26,b64) & 7.23 & 7.76 & 9.63 \\
DMFT(i26,b96) & 7.61 & 8.00 & 9.75 \\
DMFT(i26,b128) & 7.73 & 8.08 &  9.76 \\
EOM-CCSD ($3\times3\times1$) & 9.50 & 9.36 & 11.44 \\
EOM-CCSD ($6\times6\times1$) & \textbf{7.48} & \textbf{7.78} & \textbf{9.78} \\
    \hline\hline
	\end{tabular}
\end{table}

The computed direct and indirect band gaps at special $\veck$ points are presented in Table~\ref{tab:BN_gap}. In this paper, we use the notation ``DMFT(iX,bY)'' to mean that there are X impurity orbitals and Y bath orbitals to be treated by the CCSD-GF impurity solver.
The DMFT band gaps are calculated from the valence and conduction peaks of $\veck$-resolved density of states (DOS). We compare our DMFT results to HF, DFT/PBE~\cite{Perdew96PBE} and EOM-CCSD~\cite{Stanton1993,McClain2017} gaps computed using \textsc{PySCF}, all with $6\times6\times1$ $\veck$-point meshes. EOM-CCSD with $3\times3\times1$ $\veck$-point sampling is also included for comparison. As shown in Table~\ref{tab:BN_gap}, $6\times6\times1$ EOM-CCSD  with the GTH-DZVP basis predicts that 2D h-BN is a direct band-gap semiconductor, with a gap of 7.48 eV at the $\mathbf{K}$ point. The indirect band gap from $\mathbf{K}$ to $\mathbf{\Gamma}$ has a slightly larger value of 7.78 eV. These values are taken as the reference values. Compared to $6\times6\times1$ EOM-CCSD, HF overestimates the $\mathbf{K} \rightarrow \mathbf{K}$ gap by 3.8 eV, while PBE underestimates it by 2.9 eV. EOM-CCSD with a smaller $3\times3\times1$ $\veck$ mesh also overestimates the band gaps by 1.6-2.0 eV, suggesting the importance of large $\veck$-point meshes to approach the thermodynamic limit.

Even with a small number (32) of bath orbitals, our DMFT(i26,b32) result shows significant improvement over mean-field methods, although the $\mathbf{K} \rightarrow \mathbf{K}$ gap is still underestimated by 1.8 eV. As we increase the number of bath orbitals $N_b$ to 64, DMFT produces a better description of all three band gaps and the errors are all within 0.3 eV, indicating the necessity of using a sufficient number of bath orbitals to reduce the bath discretization error. Even with a  $1\times1\times1$ unit cell as the impurity, the DMFT(i26,b64) result is superior to the EOM-CCSD ($3\times3\times1$) result due to the larger amount of $\veck$-point sampling. By further increasing $N_b$ to 128, we also demonstrate that the DMFT band gaps are converged to around 0.1 eV at $N_b=96$. Thus, we believe our DMFT results for 2D h-BN to be well converged with respect to the bath size. In addition, to investigate the effect of not coupling bath orbitals to the PAO space, we further compare the IAO-only hybridization against the full IAO+PAO hybridization near the Fermi surface from the converged DMFT(i26,b96) calculation, as shown in Fig. S1 (Supporting Information). There we observe that the IAO-only hybridization captures the low-energy physics very well compared to the IAO+PAO hybridization, indicating that the error introduced by this approximation is very small. The DMFT(i26,b96) calculation takes 1.5 hours to converge on 2 nodes with 28 CPU cores per node. Further obtaining the Green's function and DOS at each frequency point takes about 4.5 minutes. This should be compared to the computational cost of a full EOM-CCSD calculation with the $6\times6\times1$ $\veck$-point mesh, which takes about 40 hours to obtain 4 IP and EA roots at all $\veck$ points.

\begin{figure}[hbt]
\includegraphics{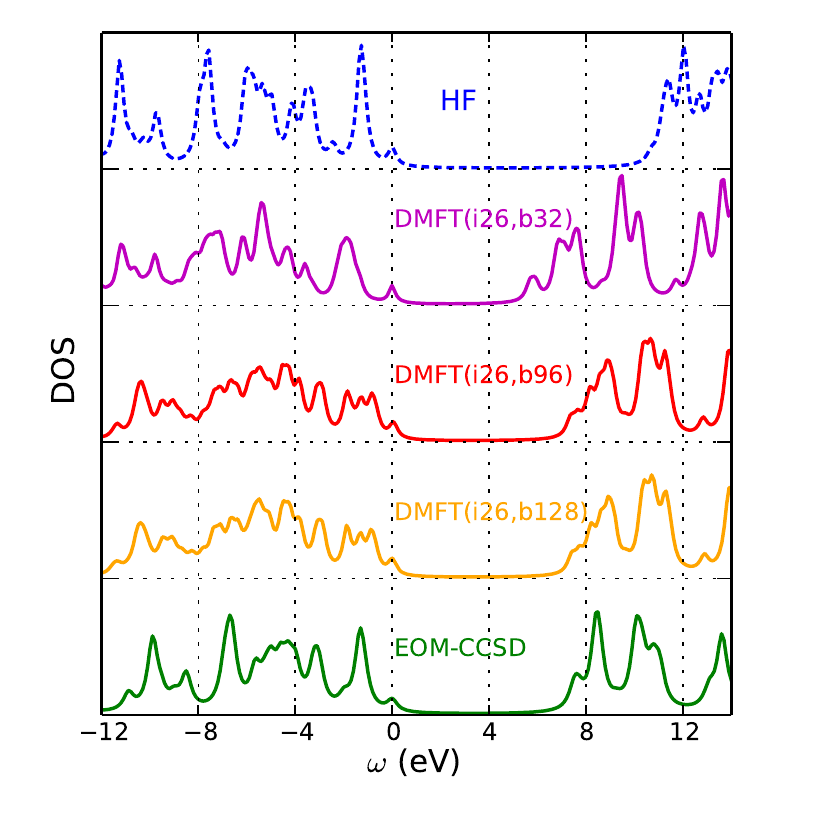}
\caption{Local density of states of 2D h-BN monolayer. DMFT spectral functions are computed with a broadening factor $\eta=0.2$ eV. EOM-CCSD ($6\times6\times1$) DOS is generated using a Lorentzian broadening.}
\label{fig:BN_DOS}
\end{figure}

We next show the local density of states (DOS) in Fig.~\ref{fig:BN_DOS}, where the DOS is computed from the spectral function: $\rho(\mathbf{R}=\mathbf{0}, \omega)=\mathrm{Tr} \mathbf{A}(\mathbf{R}=\mathbf{0},\omega)$. Here we compare DMFT with HF and $6\times6\times1$ EOM-CCSD. Since we employ only a
single $\veck$-point mesh to minimize cost, we only obtain a finite set of excitation energies from EOM-CCSD. Consequently,
we have applied a Lorentzian broadening to the IP- and EA-EOM-CCSD roots to generate the corresponding DOS spectrum,
as an approximation to the true CCSD-GF DOS in the TDL.
As can be seen in Fig.~\ref{fig:BN_DOS}, DMFT again significantly improves over HF. In addition to the much better band gaps, the DMFT DOS
also has a better structure than the HF DOS. In particular, the DMFT conduction bands are almost identical to the EOM-CCSD ones, even for the high-energy bands. This is a result of including the high-energy virtual orbitals (PAOs) into the impurity problem.  Comparing the DMFT(i26,b96) and DMFT(i26,b128) DOS plots, we find that the DMFT spectral functions are already well converged at $N_b=96$. 

\begin{figure}[hbt]
\includegraphics{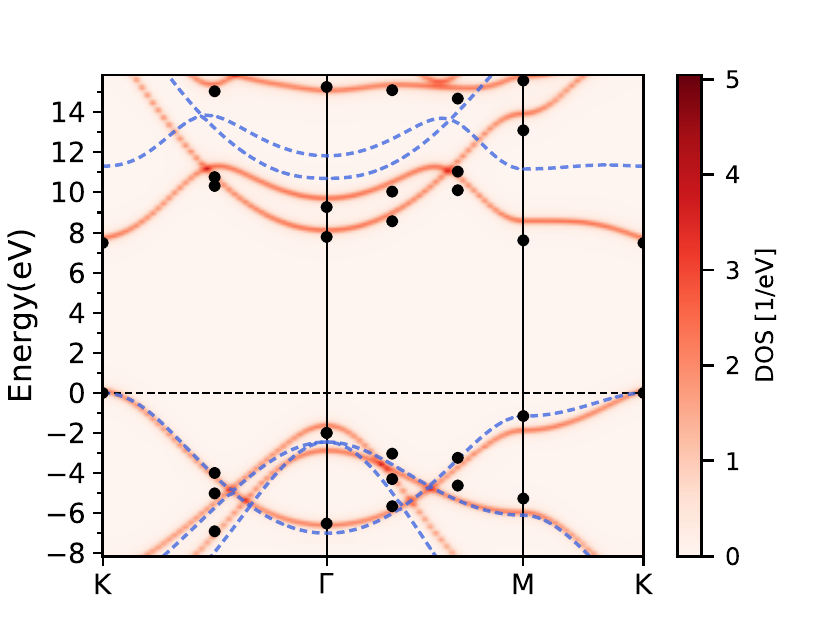}
\caption{Band structure of 2D h-BN from DMFT(i26,b96) calculation with a broadening factor $\eta=0.1$ eV (heat map). The blue dashed lines represent the HF band structure and the black circles are EOM-CCSD ($6\times6\times1$) charged excitation energies.}
\label{fig:BN_band}
\end{figure}

On the other hand, the agreement between the DMFT DOS of the valence bands and that of EOM-CCSD is less perfect.
Compared to EOM-CCSD, several valence peaks are broader in DMFT,
such as the valence peak near the Fermi surface. To understand this behavior, we plotted the DMFT band structure of h-BN using the $N_b=96$ data, as presented in Fig.~\ref{fig:BN_band}. Compared to HF, we find that the point group symmetry at certain $\veck$ points is broken in the DMFT band structure plot. For example, at the $\mathbf{\Gamma}$ point, there is a degeneracy in the highest valence band observed in both HF and EOM-CCSD, but this degeneracy is slightly broken in DMFT. This explains the broader DMFT valence bands in Fig.~\ref{fig:BN_DOS}. We believe such behavior is
due to a mismatch between the diagonal and off-diagonal parts of self-energy in DMFT: the diagonal part is computed from the impurity CCSD-GF, while the off-diagonal part is from the lattice $\veck$-point HF self-energy. Aside from the slight symmetry breaking, the DMFT band structure is in good agreement with the EOM-CCSD result. 

Overall, however, the data demonstrates that our DMFT scheme works well in 2D h-BN. The DMFT procedure produces accurate band gaps and is also
capable of modeling bands far away from the Fermi surface, even with a small number of bath orbitals ($N_b=64$).

\subsection{Bulk Si}
We next apply our DMFT implementation to the silicon crystal (Si). Silicon is a small band-gap semiconductor with delocalized valence electrons.
Such a system presents a  challenge to quantum embedding methods, including DMFT, as these methods all start from a local correlation approximation.
Here, we assess the effect of impurity size on the description of spectral functions of bulk Si. Two different impurity sizes were considered: a $1\times1\times1$ unit cell and a $2\times2\times1$ supercell. The $2\times2\times1$ cell is the largest impurity size that can currently be handled
using our CCSD-GF solver. In the $1\times1\times1$ unit cell, there are 26 impurity orbitals, 8 of which are IAOs. Two Gauss-Legendre quadratures of $N_\omega=4,20$ ($N_b=32,160$) were used to show the effect of bath size on spectral functions.
In the $2\times2\times1$ supercell impurity, there are 104 impurity orbitals.
There we used a quadrature of $N_\omega=4$ ($N_b=128$). A $2\times2\times4$ $\veck$-mesh was employed for the larger impurity to
generate a DMFT lattice $4\times4\times4$ $\veck$-mesh. We note that DMFT(i104,b128) should be directly compared to DMFT(i26,b32) to demonstrate the effect of impurity size, as these calculations share the same bath quadrature ($N_\omega=4$) and similar bath discretization error.

\begin{figure}[hbt]
\includegraphics{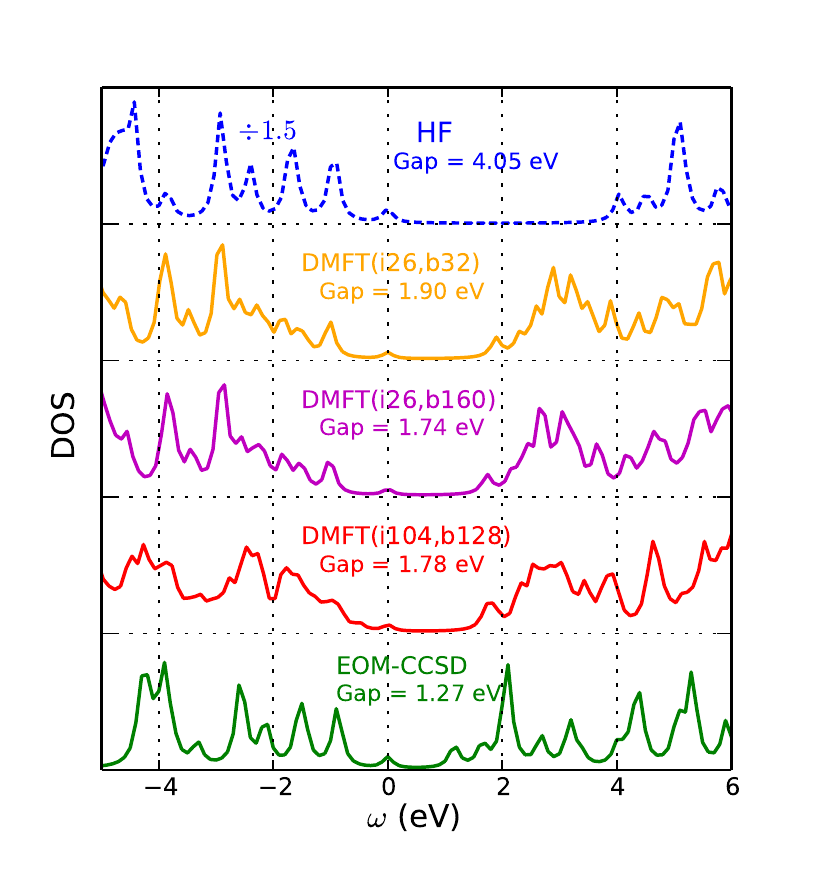}
\caption{Local density of states of bulk Si. DMFT spectral functions are computed with a broadening factor $\eta=0.1$ eV. The EOM-CCSD DOS is generated using $3\times3\times3$ $\veck$-point sampling and a Lorentzian broadening. The gap values correspond to the indirect band gap of $\mathbf{\Gamma} \rightarrow \mathbf{X}$, and the EOM-CCSD gap is from a $4\times4\times4$ $\veck$-mesh calculation.}
\label{fig:Si_DOS}
\end{figure}

The computed local DOS of bulk Si are presented in Fig.~\ref{fig:Si_DOS}. HF and EOM-CCSD results are included for comparison. A full $4\times4\times4$ $\veck$-point EOM-CCSD calculation for multiple IP/EA roots is very expensive, so we instead performed a $3\times3\times3$ $\veck$-mesh calculation. All $\veck$ points in the $4\times4\times4$ $\veck$-mesh were sampled by shifting the $3\times3\times3$ $\veck$-mesh center. The final reference DOS was then generated by applying a Lorentzian broadening. We also conducted a full $4\times4\times4$ $\veck$-mesh EOM-CCSD calculation to estimate the reference indirect  $\mathbf{\Gamma} \rightarrow \mathbf{X}$ band gap value, which we found to be 1.27 eV, as noted in Fig.~\ref{fig:Si_DOS}.

DMFT(i26,b32) produces a better $\mathbf{\Gamma} \rightarrow \mathbf{X}$ gap of 1.90 eV than HF, which overestimates the gap by 2.8 eV. Using a larger bath size of $N_b=160$ further improves the DMFT gap to be 1.74 eV.
However, the error of DMFT(i26,b160) is still around 0.5 eV, which is worse than the observed error of DMFT in 2D h-BN.
In addition, the shape of the spectrum for DMFT(i26,b160) is not very accurate.
These results support the observation that bulk Si is indeed a more difficult system for DMFT due to the stronger effects
  of the non-local interactions in such small band-gap systems.
After increasing the impurity size, we find that DMFT gives better agreement with the reference EOM-CCSD spectrum.
The DMFT(i104,b128) calculation finds the $\mathbf{\Gamma} \rightarrow \mathbf{X}$ gap to be 1.78 eV, reducing the DMFT(i26,b32) error to 0.5 eV.
This is also better than the $2\times2\times1$ $\veck$-point EOM-CCSD result, which estimates the $\mathbf{\Gamma} \rightarrow \mathbf{X}$ gap
to be 0.59 eV ($\sim$0.7 eV error). DMFT(i104,b128) does not produce a better band gap compared to DMFT(i26,b160) due to the insufficient bath size, suggesting that minimizing the bath discretization error is also important. Nevertheless, the DMFT(i104,b128) spectrum has an improved shape, especially in the valence region,
where the bands have similar peak positions to the EOM-CCSD ones. Thus, bulk Si provides a good demonstration
of the important role of impurity size in capturing the non-local self-energy in  delocalized systems.
However, even with the larger $2\times2\times1$ impurity, the DMFT results are still not completely accurate,
indicating the need for both larger impurities and a better treatment of long-range interactions than the  HF self-energy.

\subsection{NiO}

We finally turn to discuss the prototypical strongly-correlated system, NiO. NiO has a type-II AFM phase below the N\'{e}el temperature (525 K),
with ferromagnetic planes stacked in the [111] direction. Due to the partially filled $d$ orbitals in Ni, spin-unpolarized DFT methods predict NiO to be a metal, and spin-polarized DFT (LSDA/GGA) significantly underestimates the band gap and magnetic moment. DFT+DMFT simulations
  with a single Ni $3d$ impurity
have been shown to successfully reproduce features of the experimental spectral functions and band structure of NiO in the paramagnetic (PM) phase~\cite{Ren2006,Kunes2007a,Kunes2007b,Yin2008,Karolak2010,Thunstrom2012,Leonov2016}.

\begin{figure}[hbt]
\includegraphics[width=0.7\textwidth]{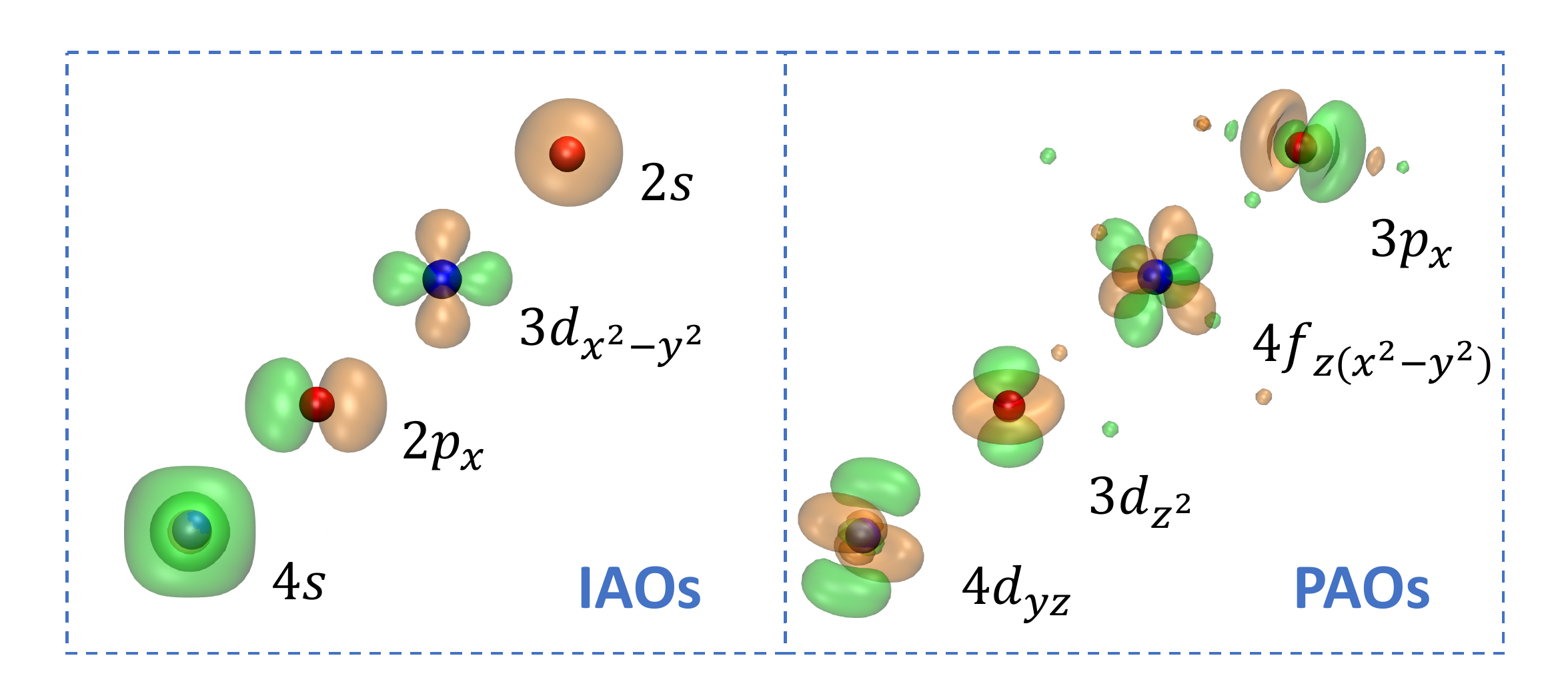}
\caption{Representative IAOs and PAOs of NiO used in DMFT calculations within a unit cell. Ni and O atoms are marked in blue and red.}
\label{fig:NiO_orbs}
\end{figure}

In this study, we use a unit cell of two Ni and two O atoms as the impurity in DMFT, corresponding to 78 impurity orbitals.
Compared to earlier single site studies, we can thus examine antiferromagnetic order within the cell, as well as the effect
of all the interactions in the crystal bands. 
Some representative IAOs and PAOs in the cell are shown in Fig.~\ref{fig:NiO_orbs}. The IAOs include the projected $3s3p3d4s$ orbitals of Ni and $2s2p$ orbitals of O, and the PAOs include the remaining $4p4d4f5s$ orbitals of Ni and $3s3p3d$ orbitals of O. Since the $3s3p$ orbitals of Ni form very flat bands far below the Fermi surface, we do not couple bath orbitals to them, to reduce the computational cost. We also used the bath truncation technique described in Sec.~\ref{sec:Hemb} to remove very weakly coupled bath orbitals, setting the eigenvalue threshold to $\lambda=0.005$ a.u. and $\lambda=0.01$ a.u. for Gauss-Legendre quadratures of $N_\omega=4$ and $N_\omega=8$ respectively. This led to a significant reduction in the number of bath orbitals, e.g., from $N_b=160$ to $N_b=86$ in the DMFT(i78,b86)@$\Phi_\mathrm{UHF}$ calculation.

\begin{figure}[hbt]
\includegraphics{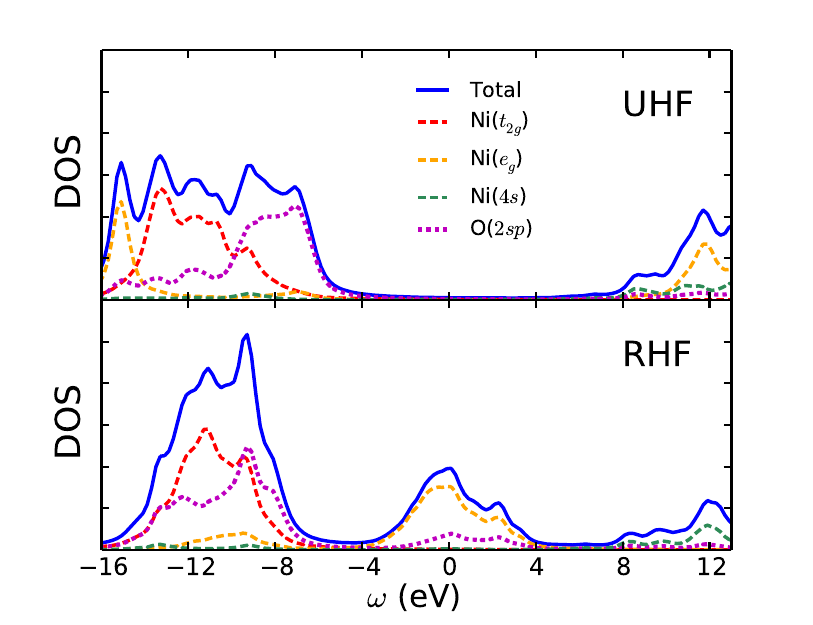}
\caption{Local density of states of NiO from UHF and RHF. A broadening factor $\eta=0.4$ eV is used.}
\label{fig:NiO_HF}
\end{figure}

To obtain an AFM solution in DMFT, we allowed spin symmetry to break by allowing different self-energies ($\vecSig_\imp^\sigma(\omega)$) and hybridizations ($\vecDel^\sigma(\omega)$) in different spin channels ($\sigma=\alpha,\beta$). The spin-unrestricted CCSD-GF (UCCSD-GF) impurity solver was employed to compute $\vecSig_\imp^\sigma(\omega)$. We started from either spin-restricted or spin-unrestricted mean-field wavefunctions to construct the embedding problem. Spin-unrestricted HF (UHF) gives an AFM solution with a large band gap for NiO, as shown in Fig.~\ref{fig:NiO_HF}. Starting from UHF, spin symmetry breaking happens already in the
initial DMFT impurity Hamiltonian ($\mathbf{H}_{\imp}$) and lattice HF self-energy. On the other hand, spin-restricted HF (RHF) with finite temperature smearing predicts NiO to be a metal (in PM phase), with no average local magnetic moment (Fig.~\ref{fig:NiO_HF}). In this case,  spin symmetry breaking is introduced \textit{only} during DMFT self-consistency, which generates symmetry-broken $\vecSig_\imp^\sigma(\omega)$ and $\vecDel^\sigma(\omega)$. 

In addition to starting from HF orbitals, we also explored the possibility of using DFT (LDA and PBE) orbitals to generate the DMFT impurity problem.
We again emphasize that only the DFT orbitals were used here and no DFT Hamiltonians or self-energies enter into our DMFT calculations.
That is to say, we still use the Fock matrix as the one-particle effective Hamiltonian, and the only difference is that the elements of  the
Fock matrix are evaluated using DFT orbitals. Since the local Fock self-energy can be exactly subtracted, there are no double counting errors in these DMFT calculations. One advantage of using DFT orbitals is that fully self-consistent HF calculations can be avoided, which is very expensive in large systems due to the quadratic scaling with respect to $\veck$ points.

\begin{table}[hbt]
	\centering
	\caption{Band gap ($E_g$), magnetic moment ($m_\mathrm{Ni}$) and local charge ($\rho_\mathrm{Ni}$) of NiO in the AFM phase. The first column of band gaps is computed from the half maxima of the first valence and conduction bands of the local DOS, as done in Ref.~\cite{Sawatzky1984}. The second column of band gaps (in  brackets) is computed from the valence and conduction peaks of the $\veck$-resolved DOS. The notation ``@$\Phi_\mathrm{HF/PBE/LDA}$'' indicates the underlying orbitals used to construct the Fock matrix in DMFT.}
	\label{tab:NiO_gap}
	\begin{tabular}{cccc}
	\hline\hline
Method	 &  $E_g$ (eV)  & $m_\mathrm{Ni}$ ($\mu_\mathrm{B}$) &  $\rho_\mathrm{Ni}$ ($e$) \\
	\hline
UHF &   - (11.6) & 1.86 & 1.42 \\
UPBE &  - (1.3) & 1.42 & 1.02 \\
ULDA &  - (0.6) & 1.28 & 0.94 \\
\hline
DMFT(i78,b50)@$\Phi_\mathrm{UHF} $ & 9.6 (9.4) & 1.80 & 1.35  \\
DMFT(i78,b86)@$\Phi_\mathrm{UHF} $ & 9.2 (9.0) & 1.80 & 1.35  \\
DMFT(i78,b52)@$\Phi_\mathrm{UPBE} $ & 7.4 (7.1) & 1.65 & 1.12  \\
DMFT(i78,b90)@$\Phi_\mathrm{UPBE} $ & 7.1 (6.5) & 1.63 & 1.11  \\
DMFT(i78,b52)@$\Phi_\mathrm{ULDA} $ & 6.5 (6.0) & 1.63 & 1.10  \\
DMFT(i78,b90)@$\Phi_\mathrm{ULDA} $ & 6.5 (5.8) & 1.60 & 1.08  \\
\hline
DMFT(i78,b56)@$\Phi_\mathrm{RHF} $ & 3.5 (3.3) & 1.67 & 1.22  \\
DMFT(i78,b98)@$\Phi_\mathrm{RHF} $ & 3.0 (3.1) & 1.60 & 1.17  \\
\hline
Exp & 4.3~\cite{Sawatzky1984} (-) & 1.77~\cite{Fender1968},1.90~\cite{Cheetham83NiO} & - \\ 
    \hline\hline
	\end{tabular}
\end{table}

We present DMFT results on the band gap, magnetic moments and local charges of NiO in Table~\ref{tab:NiO_gap}. The band gap is computed from the half maxima of the first valence and conduction bands of local DOS, as done in the XPS/BIS experiment~\cite{Sawatzky1984}. Meanwhile, we also report the band gap values  calculated from the valence and conduction peaks of the $\veck$-resolved DOS. The magnetic moments and local charges are calculated from the impurity UCCSD density matrix with atomic decomposition in the IAO+PAO basis. As can be seen, DMFT(i78,b86)@$\Phi_\mathrm{UHF}$ improves the UHF band gap by 2.6 eV and produces an accurate magnetic moment. However, the band gap is still too large when compared to experiment (4.3 eV). This is likely because the off-diagonal (inter-cell) part of the self-energy from the inaccurate initial UHF solution  has a large residual effect on the lattice Green's function.
In particular, this inaccurate off-diagonal self-energy leads to too large an amount of symmetry breaking in the initial impurity Hamiltonian, which cannot be completely corrected
by the DMFT local self-energy. 
In contrast, when using UPBE orbitals, DMFT(i78,b90)@$\Phi_\mathrm{UPBE}$ gives a better $\veck$-resolved band gap of 6.5 eV, suggesting that employing UPBE orbitals reduces the error in the off-diagonal HF self-energy. Using ULDA orbitals further improves the $\veck$-resolved band gap to 5.8 eV, although the error is still more than 1 eV. Overall, these results show that when starting from a spin-symmetry broken solution, our DMFT scheme is sensitive to the choice of underlying orbitals, which may lead to a variation of 3 eV in the predicted band gap of NiO. This sensitivity may be alleviated if charge self-consistency is further imposed outside of the DMFT loop~\cite{Granas2012,Park2014}, which is absent in our current implementation. However, the systematic overestimation of the band gap suggests that besides  charge self-consistency, non-local contributions to the self-energy need to be treated more accurately~\cite{Si1996,Sun2002,Biermann2003,Sun2004}.

When we switch to a spin-restricted HF reference, the DMFT results are more accurate than the UHF-based DMFT results. DMFT(i78,b98)@$\Phi_\mathrm{RHF}$ predicts a reasonable $\veck$-resolved band gap of 3.1 eV and magnetic moment of 1.60 $\mu_\mathrm{B}$. This superior performance can be attributed to the fact that the initial incorrect symmetry breaking in the lattice HF self-energy and impurity Hamiltonian is avoided by using a spin-restricted reference.  As a result, the spin symmetry breaking is solely determined by the accurate DMFT self-energy obtained from the
UCCSD-GF impurity solver, leading to improved results, particularly for the spectral functions. 

\begin{figure}[hbt]
\includegraphics{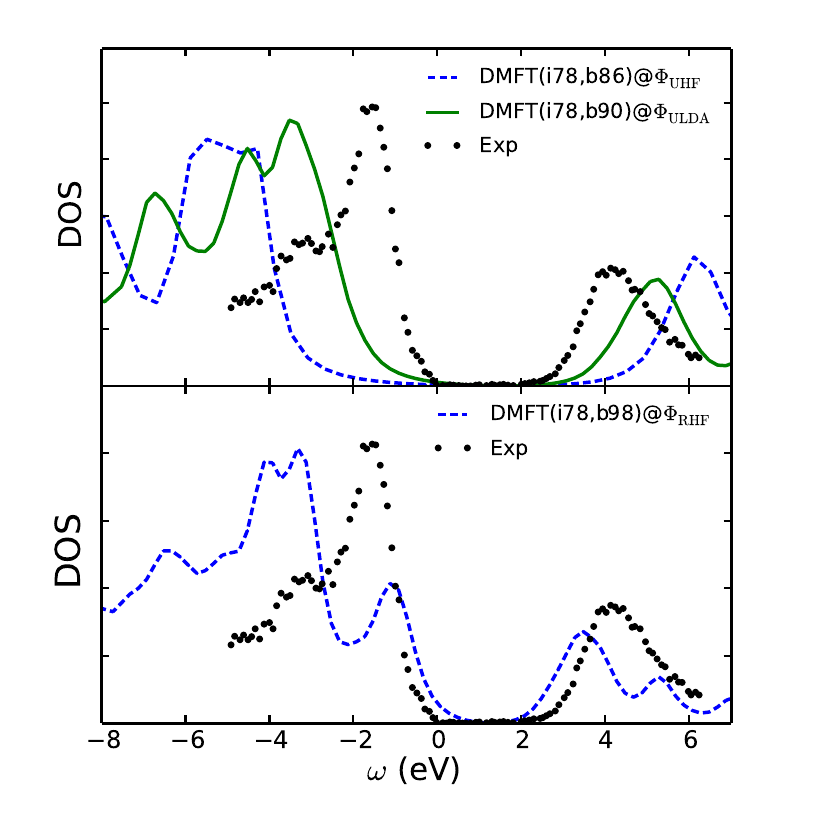}
\caption{Local density of states of NiO in the AFM phase. The XPS/BIS experimental DOS is from Ref.~\cite{Sawatzky1984}. The DMFT DOS is computed with a broadening factor $\eta=0.4$ eV, and is shifted horizontally for an easier comparison to experiment.}
\label{fig:NiO_DOS}
\end{figure}

We present the local DOS of NiO in Fig.~\ref{fig:NiO_DOS}. The upper panel shows that DMFT(i78,b86)@$\Phi_\mathrm{ULDA}$ gives a similar spectral shape
to experiment, although the band gap is larger and the first valence peak is broader. On the other hand, DMFT(i78,b90)@$\Phi_\mathrm{UHF}$ spectrum has too wide a band gap. In the lower panel, the DMFT(i78,b98)@$\Phi_\mathrm{RHF}$ result agrees well with experiment near the Fermi surface. However, the main valence peak is separated into two peaks, where the highest peak is around -3.5 eV and a shoulder peak appears around -1 eV. Such a two-peak structure is not observed in experiment. 

\begin{figure}[hbt]
\subfigure[]{\label{fig:NiO_comp:a}\includegraphics[width=0.49\textwidth]{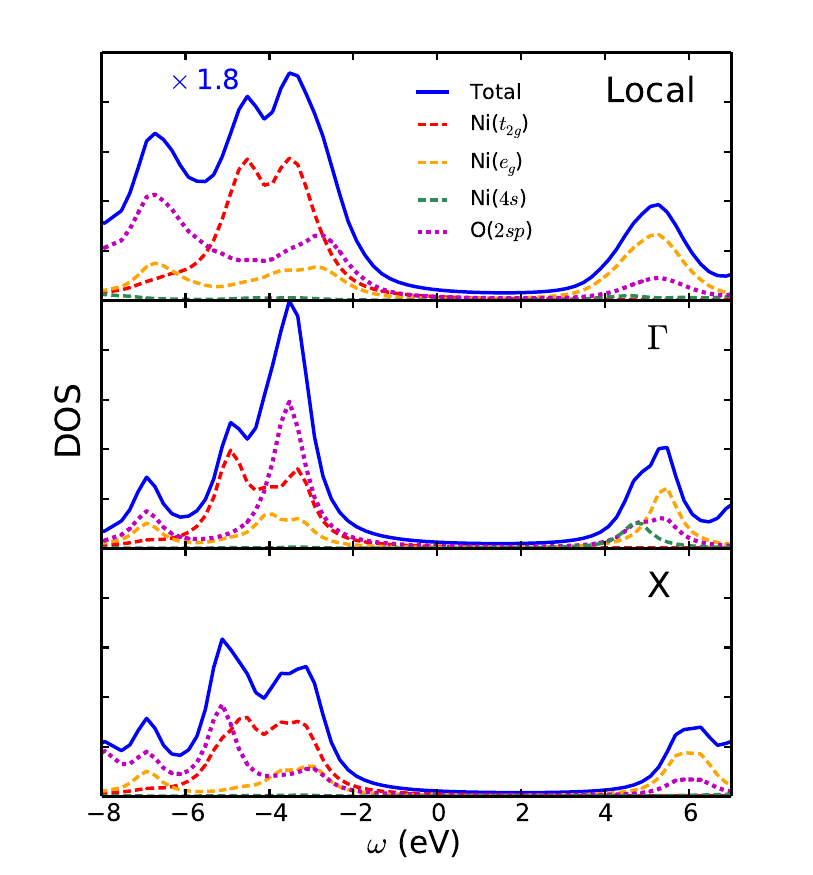}}
\subfigure[]{\label{fig:NiO_comp:b}\includegraphics[width=0.49\textwidth]{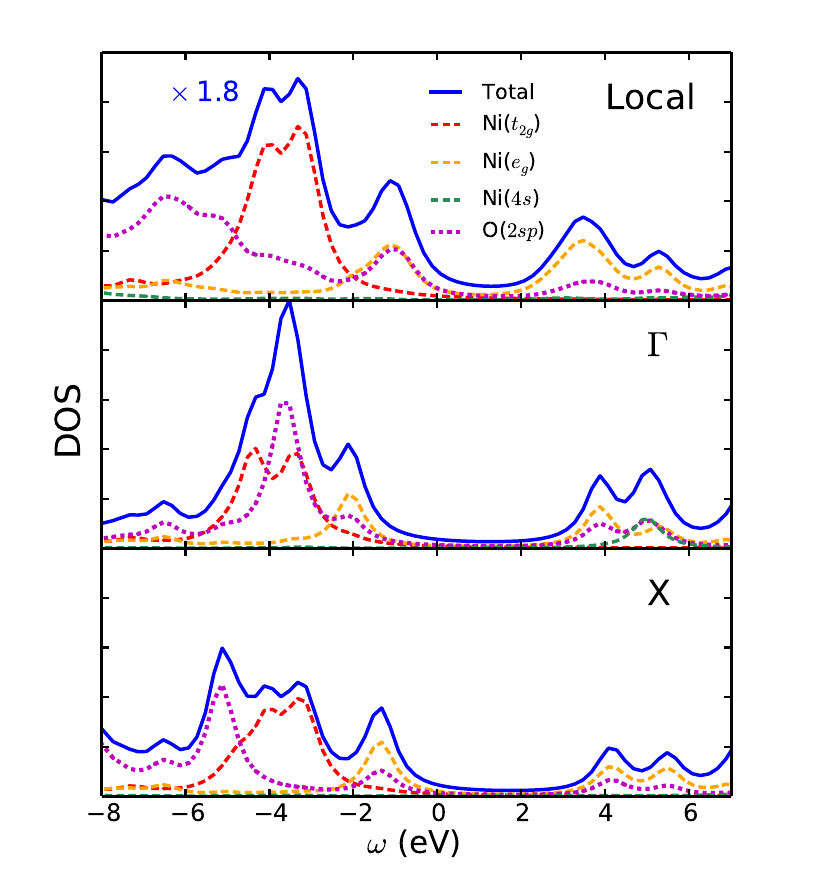}}
\caption{Components of NiO density of states from (a) DMFT(i78,b90)@$\Phi_\mathrm{ULDA}$ and (b) DMFT(i78,b98)@$\Phi_\mathrm{RHF}$ calculations. Panel 1 shows the local DOS, and Panels 2 and 3 show $\veck$-resolved DOS at $\mathbf{\Gamma}$ and $\mathbf{X}$ points.}
\label{fig:NiO_comp}
\end{figure}

To understand the deviations from experiment and to further study the character of the insulating gap, we plot the components of the NiO DOS in Fig.~\ref{fig:NiO_comp}. From Panel 1 of Fig.~\ref{fig:NiO_comp:a}, DMFT(i78,b90)@$\Phi_\mathrm{ULDA}$ predicts that the insulating gap is from a complicated charge transfer transition with mixed Mott character: the valence peak is of O $2sp$ and Ni $t_{2g}$ and $e_g$ characters, and the conduction peak is mainly of an $e_g$ character with a small O $2sp$ contribution. On the other hand, it can be seen in Fig.~\ref{fig:NiO_HF} that UHF predicts the insulating gap of NiO to be from a pure charge transfer transition from O $2p$ to Ni $e_g$. This comparison indicates that DMFT significantly corrects the positions of the Ni $t_{2g}$ and $e_g$ bands. Meanwhile, as shown in Fig.~\ref{fig:NiO_comp:b}, DMFT(i78,b98)@$\Phi_\mathrm{RHF}$ also predicts mixed charge transfer and Mott character for the insulating gap, although the first valence peak has almost no Ni $t_{2g}$ contribution. This indicates that the artificial shoulder valence peak near the Fermi surface arises because of a mismatch in the relative positions of the Ni $t_{2g}$ and O $2sp$/Ni $e_g$ bands. Once again,
these results lead us to conclude that to obtain both the band gap and spectral shape accurately, we require either larger impurities or
a more sophisticated treatment of the inter-cluster interactions.
We also plot the $\veck$-resolved DOS in Panels 2 and 3 in Fig.~\ref{fig:NiO_comp}. It can be seen that the O $2sp$ main valence peak is shifted by 2 eV at the $\mathbf{X}$ point compared to the $\mathbf{\Gamma}$ point.  Interestingly, we find that the conduction band at the $\mathbf{\Gamma}$ point has a significant contribution from the Ni $4s$ and O $2sp$ orbitals. This has not been reported in previous DFT+DMFT studies but has also been found in quasiparticle self-consistent GW calculations~\cite{Das2015} and a recent periodic EOM-CCSD calculation~\cite{Gao2019}.

\section{Conclusions}\label{sec:conclusion}
In this work, we described a new \abinitio DMFT scheme for studying realistic solid-state materials using both large impurities, spanning multiple unit cells and atoms, and realistic quantum chemical basis sets. The use of Hartree-Fock as the low-level method allowed us to avoid empirical parametrizations and double counting errors. By employing the computational infrastructure of the intrinsic atomic orbital/projected atomic orbital basis, fast integral transformations, and bath truncation, we showed that we could efficiently construct the DMFT embedding Hamiltonian. Finally, combining this scheme with coupled-cluster impurity solvers enabled us to solve the impurity problem with several hundred embedding orbitals. Our numerical results on 2D h-BN, bulk Si and NiO (in the antiferromagnetic phase) were encouraging and demonstrated the promise of \abinitio DMFT in the simulation of both weakly and strongly correlated materials.
Future work will explore the impact of charge self-consistency and a more  accurate treatment of long-range interactions.

\begin{acknowledgement}
  We thank Timothy Berkelbach and Yang Gao for helpful discussions. This work is supported by the US DOE via DE-SC0018140. Additional support
  was provided by the Simons Foundation via the Simons Collaboration on the Many-Electron Problem, and via the Simons Investigatorship in Physics.
\end{acknowledgement}

\begin{suppinfo}
Comparison of IAO-only and IAO+PAO hybridizations and causality of the CCSD self-energy in DMFT.
\end{suppinfo}

\providecommand{\latin}[1]{#1}
\makeatletter
\providecommand{\doi}
  {\begingroup\let\do\@makeother\dospecials
  \catcode`\{=1 \catcode`\}=2 \doi@aux}
\providecommand{\doi@aux}[1]{\endgroup\texttt{#1}}
\makeatother
\providecommand*\mcitethebibliography{\thebibliography}
\csname @ifundefined\endcsname{endmcitethebibliography}
  {\let\endmcitethebibliography\endthebibliography}{}

\end{document}